\documentclass[onecolumn]{pasj00}
\newcommand{\bm}{\boldmath}

\newcommand{\bml}{\mbox{\bm $l$}}

\newcommand{\bmv}{\mbox{\bm $v$}}

\newcommand{\bmnabla}{\mbox{\bm $\nabla$}}
\newcommand{\bmF}{\mbox{\bm $F$}}

\begin{document}
\SetRunningHead{J.\ Fukue}
{Radiative Transfer in Relativistic Accretion-Disk Winds}
\Received{yyyy/mm/dd}
\Accepted{yyyy/mm/dd}
%tex          2007 0616
%referee      2007 0816
%editing      2007

\title{Radiative Transfer in Relativistic Accretion-Disk Winds}

%%% begin:list of authors
\author{Jun \textsc{Fukue}} %   \thanks{}}
\affil{Astronomical Institute, Osaka Kyoiku University, 
Asahigaoka, Kashiwara, Osaka 582-8582}
\email{fukue@cc.osaka-kyoiku.ac.jp}

%\author{B-Firstname \textsc{B-Familyname}}
%\affil{B-Address of Institute}\email{bbbbb@xxx.xxx.xx.xx}
%\and
%\author{C-Firstname {\sc C-Familyname}}
%\affil{C-Address of Institute}\email{ccccc@xxx.xxx.xx.xx}
%%% end:list of authors

%% `\KeyWords{}' always has to be placed before `\maketitle'.
 %Do NOT move this preamble from here!

\maketitle

%\newpage

\begin{abstract}
Radiative transfer in a relativistic accretion disk wind 
is examined under the plane-parallel approximation
in the fully special relativistic treatment.
For an equilibrium flow, 
where the flow speed and the source function are constant,
the emergent intensity is analytically obtained.
In such an equilibrium flow
the usual limb-darkening effect does not appear,
since the source function is constant.
Due to the Doppler and aberration effects
associated with the relativistic motion of winds, however,
the emergent intensity is strongly enhanced
toward the flow direction.
This is the {\it relativistic peaking effect}.
We thus carefully treat and estimate
the appearance of relativistic winds and jets,
when we observe them in an arbitrary direction.
\end{abstract}

\section{Introduction}

Accretion disks are now widely believed
to be energy sources in various active phenomena in the universe
(see Kato et al. 2007 for a review).
Relating to energetic accretion disks,
accretion disk winds have been extensively examined
in connection with astrophysical jets and outflows:
in bipolar outflows from young stellar objects (YSOs),
in outflows from cataclysmic variables (CVs) and
supersoft X-ray sources (SSXSs),
in relativistic jets from 
microquasars ($\mu$QSOs), active galactic nuclei (AGNs), 
quasars (QSOs), and in gamma-ray bursts (GRBs).
In particular,
intense radiation fields of 
luminous supercritical accretion disks
may be responsible for relativistic jets
 from super-Eddington sources,
such as luminous $\mu$QSOs, GRS~1915$+$105 and SS~433, 
luminous QSOs, 3C~273, and energetic GRBs
(see, e.g., Fukue 2004 for references).
Furthermore,
energetic emissions from relativistic jets have been examined,
relating to, e.g., gamma-ray blazars and gamma-ray bursts
(Dermer, Schlickeiser 1993, 2002; Dermer 1998;
B\"ottcher, Dermer 2002; Dermer et al. 2007).

In such circumstances,
radiative transfer in accretion disk winds as well as
accretion disks becomes more and more important.
Radiative transfer in the standard disk has been investigated
in relation to the structure of a static disk atmosphere
and the spectral energy distribution from the disk surface
(e.g., Meyer, Meyer-Hofmeister 1982; Cannizzo, Wheeler 1984).
Furthermore, gray and non-gray models of accretion disks 
were constructed under numerical treatments
(K\v ri\v z, Hubeny 1986; Shaviv, Wehrse 1986;
Adam et al. 1988; Mineshige, Wood 1990; 
Ross et al. 1992; Shimura, Takahara 1993;
Hubeny, Hubeny 1997, 1998; Hubeny et al. 2000, 2001;
Davis et al. 2005; Hui et al. 2005)
and under analytical ones
(Hubeny 1990; Artemova et al. 1996; Fukue, Akizuki 2006a).

Radiative transfer in the accretion disk wind, on the other hand,
has not been well considered
both in the non-relativistic and relativistic regimes.
For example,
transformation properties of disk radiation fields
in the proper frame of a relativistic jet
were examined by, e.g., Dermer and Schlickeiser (2002).
In these earlier works, however,
the radiation fields are set to be external sources,
and the radiation transfer was not considered.
Recently,
radiative transfer in a moving disk atmosphere
was firstly investigated
in the subrelativistic regime
(Fukue 2005a, 2006a),
and in the relativistic regime
(Fukue 2005b, 2006b; Fukue, Akizuki 2006b).
In these studies, however, 
only the radiative moments were obtained
under the moment formalism,
and the specific intensity was not solved.
Hence,
in Fukue (2007)
the specific intensity from an accretion disk wind
was obtained in the subrelativistic regime,
where the flow speed $v$ is of the order of $(v/c)^1$.

In this paper,
we thus extend the previous work.
Namely, 
we examine radiative transfer in the relativistic accretion disk wind,
which is assumed to blow off from the luminous disk
in the vertical direction (plane-parallel approximation),
with the relativistic speed up to the order of $c$.

%In contrast to the static atmosphere,
%in the moving atmosphere
%the boundary condition at the surface of zero optical depth
%should be modified (Fukue 2005a, b).
%Moreover,
%the usual Eddington approximation violates
%in the highly relativistic flow (Fukue 2005b;
%see also 
%Turolla, Nobili 1988; Nobili et al. 1991;
%Turolla et al. 1995; Dullemond 1999),
%and the velocity-dependent variable Eddington factor
%was proposed (Fukue 2006b for a plane-parallel case;
%Akizuki, Fukue 2007 for a spherical case).

%Radiation hydrodynamical (RHD) simulations were also performed
%for radiation-dominated supercritical disks with winds
%by several researchers
%(Eggum et al. 1985, 1988; Okuda et al. 1997, 2005; 
%Okuda, Fujita 2000; Okuda 2002;
%Ohsuga et al. 2005; Ohsuga 2006).
%In these current studies of RHD simulations for disks and winds,
%they were done in the subrelativistic regime
%up to the order of $(v/c)^1$, using the moment formalism
%and the flux-limited diffusion (FLD) approximation
%(Levermore, Pomraning 1981).
%The flux-limited diffusion method provides
%good approximations to the exact solutions
%but only if they are derived from transfer equations
%in which terms of the order of $(v/c)^2$ or higher have been retained
%(Yin, Miller 1995).

In the next section
we describe the basic equations.
In section 3, we show analytical solutions
of the specific intensity.
The final section is devoted to concluding remarks.

%%%%%%%%%%%%%%%%%%%%%%%%%%%%%%%%%%%%%%%%%%

\section{Relativistic Radiative Transfer Equation}

Let us suppose a luminous flat disk,
deep inside which
gravitational or nuclear energy is released
via viscous heating or other processes.
The radiation energy is transported in the vertical direction,
and the disk gas, itself, also moves in the vertical direction
as a {\it disk wind}
due to the action of radiation pressure (i.e., plane-parallel approximation).
For simplicity, in the present paper,
the radiation field is considered to be sufficiently intense that
both the gravitational field of, e.g., the central object
and the gas pressure can be ignored.
We also assume the gray approximation,
where the opacities do not depend on the frequency.
As for the order of the flow velocity $v$,
we consider the fully special relativistic regime.

The radiative transfer equations 
are given in several literatures
(Chandrasekhar 1960; Mihalas 1970; Rybicki, Lightman 1979;
Mihalas, Mihalas 1984; Shu 1991; Kato et al. 1998, 2007;
Peraiah 2002; Castor 2004).
The radiative transfer equation in the fully relativistic form
is given in, e.g., the appendix E of Kato et al. (2007)
in general and vertical forms.

In a general form
the radiative transfer equation in the inertial (fixed) frame
is expressed as
\begin{eqnarray}
%\lefteqn{
   \frac{1}{c} \frac{\partial I}{\partial t} + 
   \left( \bml \cdot \bmnabla \right) I
    &=&  \rho \gamma^{-3} \left( 1- \frac{\bmv\cdot\bml}{c} \right)^{-3}
%}
         \left[
            \frac{j_0}{4\pi}
            -\left(\kappa_0^{\rm abs}+\kappa_0^{\rm sca}\right)
             \gamma^{4} \left( 1- \frac{\bmv\cdot\bml}{c} \right)^{4} I
         \right.
\nonumber \\
   &&       
\hspace{2cm}
            + \frac{\kappa_0^{\rm sca} }{4\pi} \frac{3}{4} 
              \gamma^{-2} \left( 1- \frac{\bmv\cdot\bml}{c} \right)^{-2}
         \left\{
         \gamma^4 \left[ \left( 1- \frac{\bmv\cdot\bml}{c} \right)^2
         + \left( \frac{v^2}{c^2} -\frac{\bmv\cdot\bml}{c} \right)^2 \right] cE
%\nonumber \\
         \right.
\nonumber \\
   &&  
\hspace{2cm}
        + 2\gamma^2 \left( \frac{v^2}{c^2} -\frac{\bmv\cdot\bml}{c} \right)
           {\bmF\cdot\bml}
         -2\gamma^4 \left[
          \left( 1- \frac{\bmv\cdot\bml}{c} \right)^2
          + \left( 1- \frac{\bmv\cdot\bml}{c} \right)
            \left( \frac{v^2}{c^2} -\frac{\bmv\cdot\bml}{c} \right) \right]
            \frac{\bmv\cdot\bmF}{c}
\nonumber \\
   &&  
\hspace{2cm}
         \left. \left.
         + l_{i} l_{j} cP^{ij}
         - 2\gamma^2 \left( 1- \frac{\bmv\cdot\bml}{c} \right)
           {v_i l_j P^{ij}}
         + 2\gamma^4 \left( 1- \frac{\bmv\cdot\bml}{c} \right)^2
           \frac{v_i v_j P^{ij}}{c}
         \right\}
         \right],
\label{itransf}
\end{eqnarray}
Here, $\bmv$ is the flow velocity, $c$ is the speed of light, and
$\gamma$ ($=1/\sqrt{1-v^2/c^2}$) is the Lorentz factor.
In the left-hand side
the frequency-integrated specific intensity $I$ and
the direction cosine $\bml$ are quantities
measured in the inertial (fixed) frame.
In the right-hand side, 
the mass density $\rho$,
the frequency-integrated mass emissivity $j_0$, 
the frequency-integrated mass absorption coefficient $\kappa_0^{\rm abs}$,
and
the frequency-integrated mass scattering coefficient $\kappa_0^{\rm sca}$
are quantities measured in the comoving (fluid) frame,
whereas
the frequency-integrated radiation energy density $E$,
the frequency-integrated radiative flux $\bmF$, and
the frequency-integrated radiation stress tensor $P^{ij}$
are quantities measured in the inertial (fixed) frame.

In the plane-parallel geometry with the vertical axis $z$
and the direction cosine $\mu$ ($=\cos \theta$),
the transfer equation is expressed as
\begin{eqnarray}
   \mu \frac{dI}{dz}
    &=&  \rho \frac{1}{\gamma^3 (1-\beta \mu)^3}
         \left[
            \frac{j_0}{4\pi}
            -\left(\kappa_0^{\rm abs}+\kappa_0^{\rm sca}\right)
             \gamma^{4} \left( 1-\beta \mu \right)^{4} I
         \right.
          + \frac{\kappa_0^{\rm sca} }{4\pi} \frac{3}{4} \gamma^2 
         \left\{ 
           \left[ 1+\frac{(\mu-\beta)^2}{(1-\beta \mu)^2}\beta^2
                   +\frac{(1-\beta^2)^2}{(1-\beta \mu)^2}
                    \frac{1-\mu^2}{2} \right] cE
         \right.
\nonumber \\
   &&
\hspace{3cm}
         - \left[ 1+ \frac{(\mu-\beta)^2}{(1-\beta \mu)^2} \right] 2F \beta
         \left. \left.
         + \left[ \beta^2 + \frac{(\mu-\beta)^2}{(1-\beta \mu)^2}
                          - \frac{(1-\beta^2)^2}{(1-\beta \mu)^2}
                            \frac{1-\mu^2}{2} \right] cP
         \right\}
         \right],
\label{itransf_pp}
\end{eqnarray}
where $\beta$ ($=v/c$) is the normalized vertical speed, and
$F$ and $P$ are the vertical component
of the radiative flux and the radiation stress tensor
measured in the inertial frame, respectively.

For the convenience of readers,
we shall show the full set of radiation hydrodynamical equations
under the plane-parallel approximation,
although we do not use all of them in this paper.

For matter, 
the continuity equation, the equation of motion, and the energy equation
become, respectively,
\begin{eqnarray}
   \rho cu &=& \rho \gamma \beta c = J ~(={\rm const.}),
\label{continuity_pp}
\\
   c^2u\frac{du}{dz} &=& c^2 \gamma^4 \beta \frac{d\beta}{dz}
                      = -\frac{d\psi}{dz} 
               - \gamma^2 \frac{c^2}{\varepsilon + p}\frac{dp}{dz}
              +\frac{\rho c^2}{\varepsilon + p}
                   \frac{\kappa_0^{\rm abs}+\kappa_0^{\rm sca}}{c} \gamma^3 
                    \left[ \frac{}{}
                F (1+\beta^2) - (cE+cP)\beta \frac{}{} \right],
\label{motion_pp}
\\
   0 &=& \frac{q^+}{\rho} - \left( \frac{}{} j_0 - \kappa_0^{\rm abs} cE \gamma^2 
                  - \kappa_0^{\rm abs} cP u^2
                  + 2 \kappa_0^{\rm abs} F \gamma u \frac{}{} \right),
\label{energy_pp}
\end{eqnarray}
where $u$ ($=\gamma \beta$) is the vertical four velocity, 
$J$ the mass-loss rate per unit area,
$\psi$ the gravitational potential, 
$\varepsilon$ the internal energy per unit proper volume,
$p$ the gas pressure, and
$q^+$ the internal heating.
In the energy equation (\ref{energy_pp})
the advection terms in the left-hand side
are dropped under the present cold approximation.

For radiation,
the zeroth and first moment equations,
and the closure relation become, respectively,
\begin{eqnarray}
   \frac{dF}{dz} &=& \rho \gamma
         \left[ \frac{}{} j_0 - \kappa_0^{\rm abs} cE
                 + \kappa_0^{\rm sca} (cE+cP) \gamma^2 \beta^2  
        + \kappa_0^{\rm abs} F \beta
               - \kappa_0^{\rm sca} F ( 1+ \beta^2 )\gamma^2 \beta \frac{}{} \right].
\label{mom0_pp}
\\
   \frac{dP}{dz} &=& \frac{\rho \gamma}{c} 
         \left[ \frac{}{} j_0 \beta - \kappa_0^{\rm abs}  F
                  + \kappa_0^{\rm abs} cP \beta 
      - \kappa_0^{\rm sca} F \gamma^2 (1+\beta^2)
               + \kappa_0^{\rm sca} (cE+cP)\gamma^2 \beta \frac{}{} \right],
\label{mom1_pp}
\\
   cP(1-f\beta^2) &=& cE(f-\beta^2) + 2F\beta(1-f),
\label{closure_pp}
\end{eqnarray}
where 
$f(\tau, \beta)$ is the variable Eddington factor,
which is defined by $f=P_{\rm co}/E_{\rm co}$,
$E_{\rm co}$ and $P_{\rm co}$ being the comoving quantities,
and generally depends on the velocity 
and its gradient as well as the optical depth (Fukue 2006b, 2007b).

Eliminating $j_0$ using the energy equation (\ref{energy_pp}),
the transfer equation (\ref{itransf_pp}) becomes
\begin{eqnarray}
   \mu \frac{dI}{dz}
    &=&  \rho \frac{1}{\gamma^3 (1-\beta \mu)^3}
         \left[
            -\left(\kappa_0^{\rm abs}+\kappa_0^{\rm sca}\right)
             \gamma^{4} \left( 1-\beta \mu \right)^{4} I
          +  \frac{q^+}{4\pi \rho}
          + \frac{\kappa_0^{\rm abs} }{4\pi} \gamma^2 
            (cE - 2F\beta + \beta^2 cP)
         \right.
\nonumber \\
    &&
\hspace{3cm}
          + \frac{\kappa_0^{\rm sca} }{4\pi} \frac{3}{4} \gamma^2 
         \left\{ 
           \left[ 1+\frac{(\mu-\beta)^2}{(1-\beta \mu)^2}\beta^2
                   +\frac{(1-\beta^2)^2}{(1-\beta \mu)^2}
                    \frac{1-\mu^2}{2} \right] cE
         \right.
\nonumber \\
   &&
\hspace{5cm}
         - \left[ 1+ \frac{(\mu-\beta)^2}{(1-\beta \mu)^2} \right] 2F \beta
         \left. \left.
         + \left[ \beta^2 + \frac{(\mu-\beta)^2}{(1-\beta \mu)^2}
                          - \frac{(1-\beta^2)^2}{(1-\beta \mu)^2}
                            \frac{1-\mu^2}{2} \right] cP
         \right\}
         \right].
\label{itransf_pp2}
\end{eqnarray}

Introducing the optical depth defined by
\begin{equation}
   d\tau = - \left(\kappa_0^{\rm abs}+\kappa_0^{\rm sca}\right) \rho dz,
\end{equation}
the transfer equation (\ref{itransf_pp2}) finally becomes
\begin{eqnarray}
   \mu \frac{dI}{d\tau}
    &=&  \frac{1}{\gamma^3 (1-\beta \mu)^3}
         \left[
             \gamma^{4} \left( 1-\beta \mu \right)^{4} I
          - \frac{q^+}{4\pi \left(\kappa_0^{\rm abs}+\kappa_0^{\rm sca}\right)\rho}
          - \frac{1}{4\pi}
            \frac{\kappa_0^{\rm abs} }{\kappa_0^{\rm abs}+\kappa_0^{\rm sca}}
            \gamma^2 
            (cE - 2F\beta + \beta^2 cP)
         \right.
\nonumber \\
    &&
\hspace{3cm}
          - \frac{1}{4\pi}
         \frac{\kappa_0^{\rm sca} }{\kappa_0^{\rm abs}+\kappa_0^{\rm sca}}
         \frac{3}{4} \gamma^2 
         \left\{ 
           \left[ 1+\frac{(\mu-\beta)^2}{(1-\beta \mu)^2}\beta^2
                   +\frac{(1-\beta^2)^2}{(1-\beta \mu)^2}
                    \frac{1-\mu^2}{2} \right] cE
         \right.
\nonumber \\
   &&
\hspace{5cm}
         - \left[ 1+ \frac{(\mu-\beta)^2}{(1-\beta \mu)^2} \right] 2F \beta
         \left. \left.
         + \left[ \beta^2 + \frac{(\mu-\beta)^2}{(1-\beta \mu)^2}
                          - \frac{(1-\beta^2)^2}{(1-\beta \mu)^2}
                            \frac{1-\mu^2}{2} \right] cP
         \right\}
         \right].
\label{itransf_pp3}
\end{eqnarray}

Similarly, radiation hydrodynamical equations 
(\ref{motion_pp}), (\ref{mom0_pp}), and (\ref{mom1_pp}),
with the help of continuity equation (\ref{continuity_pp})
and the closure relation (\ref{closure_pp}),
become (cf. Fukue 2005b, 2006b)
\begin{eqnarray}
   c^2 J \gamma^3 \frac{d\beta}{d\tau} 
      &=& - \gamma^3 [ F(1+\beta^2) - (cE+cP) \beta]
       =  - \gamma \frac{F(f+\beta^2) -cP(1+f)\beta}{f-\beta^2},
\label{motion_pp3}
\\
   \frac{dF}{d\tau}
      &=& - \frac{q^+ \gamma}{\left(\kappa_0^{\rm abs}+\kappa_0^{\rm sca}\right)\rho}
          + \gamma^3 \beta [ F(1+\beta^2) - (cE+cP) \beta]
       =  - \frac{q^+ \gamma}{\left(\kappa_0^{\rm abs}+\kappa_0^{\rm sca}\right)\rho}
          + \gamma \beta \frac{F(f+\beta^2) -cP(1+f)\beta}{f-\beta^2},
\label{mom0_pp3}
\\
   \frac{dP}{d\tau}
      &=& - \frac{q^+ u}{\left(\kappa_0^{\rm abs}+\kappa_0^{\rm sca}\right)\rho}
          + \gamma^3 [ F(1+\beta^2) - (cE+cP) \beta]
       =  - \frac{q^+ u}{\left(\kappa_0^{\rm abs}+\kappa_0^{\rm sca}\right)\rho}
          + \gamma \frac{F(f+\beta^2) -cP(1+f)\beta}{f-\beta^2}.
\label{mom1_pp3}
\end{eqnarray}
Here, we dropped the gravitational and pressure forces.
For such a radiation-dominated flow,
where the gravitational and pressure forces are neglected,
there are two integrals (Fukue 2005b, 2006b):
\begin{eqnarray}
   c^2 J \gamma +F &=& c^2J + F_0
                 - \int_{\tau_0}^\tau
         \frac{q^+}{\left(\kappa_0^{\rm abs}+\kappa_0^{\rm sca}\right)\rho}
         \gamma d\tau,
\label{F_pp}
\\
   c^2Ju + cP &=& cP_0
                 - \int_{\tau_0}^\tau
         \frac{q^+}{\left(\kappa_0^{\rm abs}+\kappa_0^{\rm sca}\right)\rho}
         u d\tau.
\label{P_pp}
\end{eqnarray}

\section{Analytical Solutions}

In order to solve the transfer equation (\ref{itransf_pp3}) analytically,
we suppose several assumptions;
we assume that there is no internal heating ($q^+=0$) and
the flow reaches the equilibrium state,
where the flow speed is almost constant ($\beta=$ const.).
In this case, from equations (\ref{F_pp}) and (\ref{P_pp}), and
the closure relation (\ref{closure_pp}),
the radiative flux $F$, the radiation stress tensor $P$,
and the radiation energy density $E$ are all constant.
We further assume that the disk has a finite optical depth, and
there exists a uniform isotropic source of intensity $I_0$
at the disk equator of optical depth $\tau_0$.

Now, it is not so difficult to integrate
the transfer equation (\ref{itransf_pp3}).
We first rewrite equation (\ref{itransf_pp3}) symbolically as
\begin{equation}
   \mu \frac{dI}{d\tau}
    =  \gamma \left( 1-\beta \mu \right) I - S_{\rm a}' - S_{\rm s}',
\end{equation}
where
\begin{eqnarray}
   S_{\rm a}' &=& \frac{S_{\rm a}}{ \gamma^3 \left( 1-\beta \mu \right)^3 }
     =     \frac{1}{4\pi}
           \frac{\kappa_0^{\rm abs} }{\kappa_0^{\rm abs}+\kappa_0^{\rm sca}}
           \frac{1}{ \gamma \left( 1-\beta \mu \right)^3 }
            (cE - 2F\beta + \beta^2 cP)
\label{Sa_pp}
\\
   S_{\rm s}' &=& \frac{S_{\rm s}}{ \gamma^3 \left( 1-\beta \mu \right)^3 }
     =     \frac{1}{4\pi}
           \frac{\kappa_0^{\rm sca} }{\kappa_0^{\rm abs}+\kappa_0^{\rm sca}}
           \frac{3}{4}
           \frac{1}{ \gamma \left( 1-\beta \mu \right)^3 } 
         \left\{ 
           \left[ 1+\frac{(\mu-\beta)^2}{(1-\beta \mu)^2}\beta^2
                   +\frac{(1-\beta^2)^2}{(1-\beta \mu)^2}
                    \frac{1-\mu^2}{2} \right] cE
         \right.
\nonumber \\
   &&
\hspace{5cm}
         - \left[ 1+ \frac{(\mu-\beta)^2}{(1-\beta \mu)^2} \right] 2F \beta
         \left.
         + \left[ \beta^2 + \frac{(\mu-\beta)^2}{(1-\beta \mu)^2}
                          - \frac{(1-\beta^2)^2}{(1-\beta \mu)^2}
                            \frac{1-\mu^2}{2} \right] cP
         \right\},
\label{Ss_pp}
\end{eqnarray}
where $S_{\rm a}' + S_{\rm s}'$
is the Doppler boosted source function and
$S_{\rm a} + S_{\rm s}$ is the non-boosted source function,
both being independent of the optical depth under the present approximation.

Under the above situations,
we can formally integrate the transfer equation (\ref{itransf_pp3}),
similar to Fukue and Akizuki (2006) and Fukue (2007).
After several partial integrations,
we obtain both an outward intensity $I(\tau, \mu, \beta)$ ($\mu>0$)
and an inward intensity $I(\tau, -\mu, \beta)$ as
\begin{eqnarray}
   I(\tau, \mu, \beta) &=& 
         \frac{S_{\rm a}+S_{\rm s}}{ \gamma^4 \left( 1-\beta \mu \right)^4 }
         \left[
                1 - e^{\displaystyle 
                       \frac{\displaystyle \gamma(1-\beta\mu)}
                            {\displaystyle \mu}(\tau -\tau_0)}
         \right]
        + I(\tau_0, \mu) e^{\displaystyle 
                            \frac{\displaystyle \gamma(1-\beta\mu)}
                                 {\displaystyle \mu}(\tau -\tau_0)},
\label{i_sol11} \\
   I(\tau, -\mu, \beta) &=&
         \frac{S_{\rm a}+S_{\rm s}}{ \gamma^4 \left( 1-\beta \mu \right)^4 }
         \left[
                1 - e^{\displaystyle 
                       -\frac{\displaystyle \gamma(1-\beta\mu)}
                             {\displaystyle \mu}\tau}
         \right],
\label{i_sol12}
\end{eqnarray}
where $I(\tau_0, \mu)$ is the boundary value
at the wind base on the luminous disk.
These analytical solutions are reduced to
those obtained in Fukue (2007)
in the subrelativistic limit of $\gamma=1$.

In general case with finite optical depth $\tau_0$
and uniform incident intensity $I_0$ from the disk,
the boundary value $I(\tau_0, \mu, \beta)$ of the outward intensity $I$
consists of two parts:
\begin{equation}
   I(\tau_0, \mu, \beta) = I_0 + I(\tau_0, -\mu, \beta),
\end{equation}
where $I_0$ is the uniform incident intensity and
$I(\tau_0, -\mu, \beta)$ is the {\it inward} intensity from
the backside of the disk beyond the midplane.
Determining $I(\tau_0, -\mu, \beta)$ from equation (\ref{i_sol12}),
we finally obtain the outward intensity as
\begin{eqnarray}
   I(\tau, \mu, \beta) &=& 
         \frac{S_{\rm a}+S_{\rm s}}{ \gamma^4 \left( 1-\beta \mu \right)^4 }
         \left[
                1 - e^{\displaystyle 
                       \frac{\displaystyle \gamma(1-\beta\mu)}
                            {\displaystyle \mu}(\tau -2\tau_0)}
         \right]
        + I_0 e^{\displaystyle 
                            \frac{\displaystyle \gamma(1-\beta\mu)}
                                 {\displaystyle \mu}(\tau -\tau_0)}.
\label{i_sol13}
\end{eqnarray}

Finally, the emergent intensity $I(0, \mu, \beta)$ emitted from the wind top
becomes
\begin{eqnarray}
   I(0, \mu, \beta) &=& 
         \frac{S_{\rm a}+S_{\rm s}}{ \gamma^4 \left( 1-\beta \mu \right)^4 }
         \left[
                1 - e^{\displaystyle 
                       - \frac{\displaystyle \gamma(1-\beta\mu)}
                              {\displaystyle \mu} 2\tau_0}
         \right]
        + I_0 e^{\displaystyle 
                       - \frac{\displaystyle \gamma(1-\beta\mu)}
                              {\displaystyle \mu} \tau_0},
\nonumber \\
   & \sim &
         \frac{S_{\rm a}+S_{\rm s}}{ \gamma^4 \left( 1-\beta \mu \right)^4 }
     ~~~~~{\rm for~large~}\tau_0.
\label{i_sol10}
\end{eqnarray}

In order to calculate the source functions, $S_{\rm a}$ and $S_{\rm s}$,
we consider two special cases below:
a terminal case and an optically thin limit.

\subsection{Terminal Case}

When the flow speed is almost the equilibrium one
and the radiation field is almost constant,
then the values of the quantities of radiation fields
are almost equal to those at the flow top.

At the flow top of a moving photosphere at a relativistic speed,
the usual boundary conditions for a static atmosphere is inadequate,
as already pointed out in Fukue (2005b)
Namely,
the radiation field just above the wind top changes
when the gas itself does move upward,
since the direction and intensity of radiation
change due to the relativistic aberration and Doppler effect
(cf. Kato et al. 1998, 2007; Fukue 2000).
If a flat infinite plane with surface intensity $I_{\rm s}$
in the comoving frame is not static,
but moving upward at a speed $v_{\rm s}$ 
($=c\beta_{\rm s}$, and
the corresponding Lorentz factor is $\gamma_{\rm s}$),
where the subscript s denotes the values at the surface,
then, just above the surface,
the radiation energy density $E_{\rm s}$, 
the radiative flux $F_{\rm s}$, and
the radiation pressure $P_{\rm s}$ measured in the inertial frame
become, respectively,
\begin{eqnarray}
   cE_{\rm s} 
   &=& {2\pi I_{\rm s}\gamma_{\rm s}^2}
       \frac{3+3\beta_{\rm s}+\beta_{\rm s}^2}{3},
\label{Es2}
\\
   F_{\rm s}
   &=& {2\pi I_{\rm s}\gamma_{\rm s}^2}
       \frac{3+8\beta_{\rm s}+3\beta_{\rm s}^2}{6},
\label{Fs2}
\\
   cP_{\rm s}
   &=& {2\pi I_{\rm s}\gamma_{\rm s}^2}
       \frac{1+3\beta_{\rm s}+3\beta_{\rm s}^2}{3}.
\label{Ps2}
\end{eqnarray}

In this case
the non-boosted source functions are calculated as
\begin{equation}
   S_{\rm a}
     =     \frac{\kappa_0^{\rm abs} }{\kappa_0^{\rm abs}+\kappa_0^{\rm sca}}
           \frac{I_{\rm s}}{2},
~~~~~
   S_{\rm s}
     =     \frac{\kappa_0^{\rm sca} }{\kappa_0^{\rm abs}+\kappa_0^{\rm sca}}
           \frac{I_{\rm s}}{2};
~~~~~
   S_{\rm a} + S_{\rm s}
     =     \frac{I_{\rm s}}{2},
\end{equation}
which depend on neither the flow speed nor direction cosine.
Even if the non-boosted source function is constant,
the emergent intensity (\ref{i_sol10})
does depend on the flow speed and the direction cosine.
Examples of solutions are shown in figures 1 and 2.

\begin{figure}
  \begin{center}
  \FigureFile(80mm,80mm){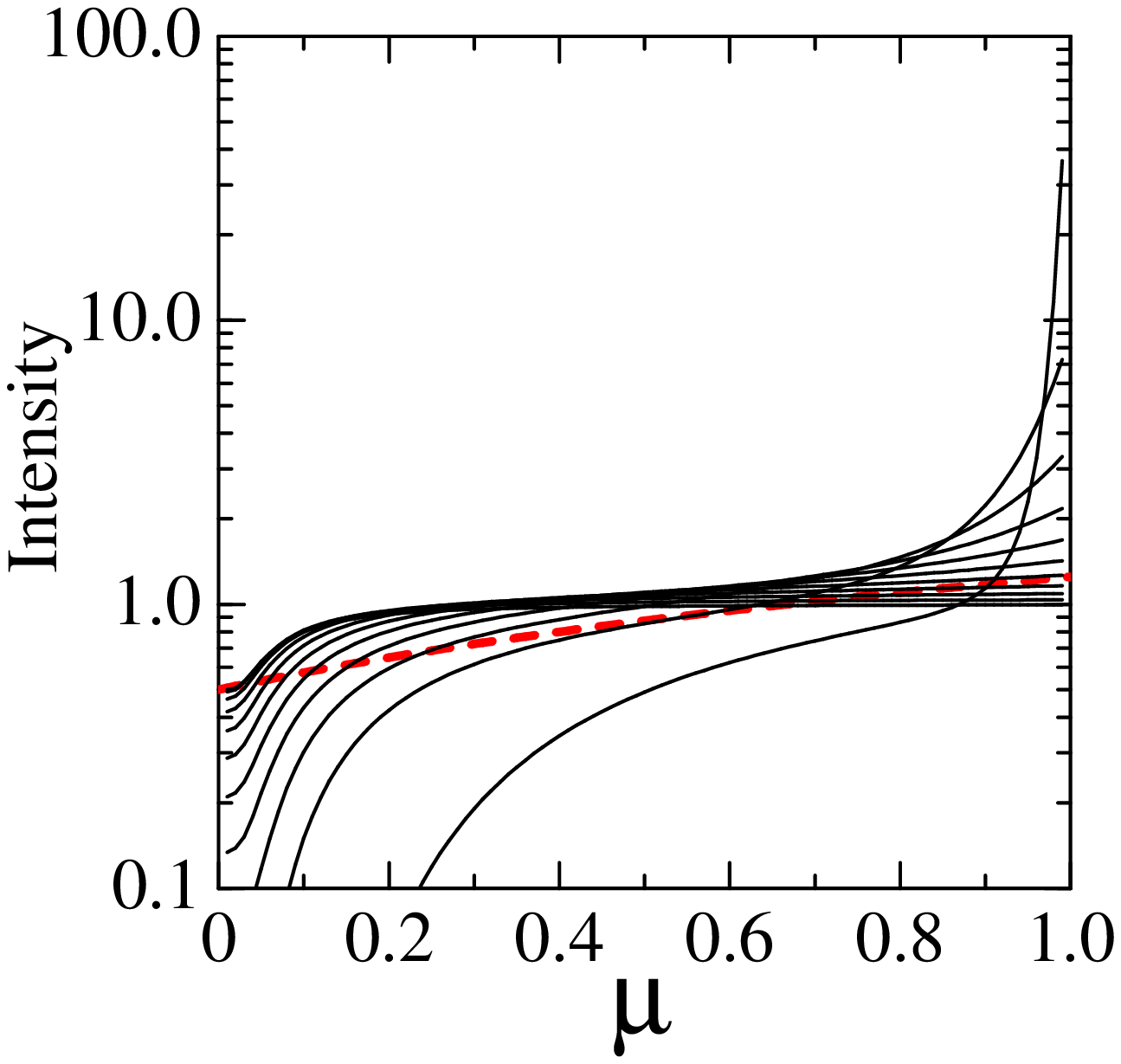}
  \FigureFile(80mm,80mm){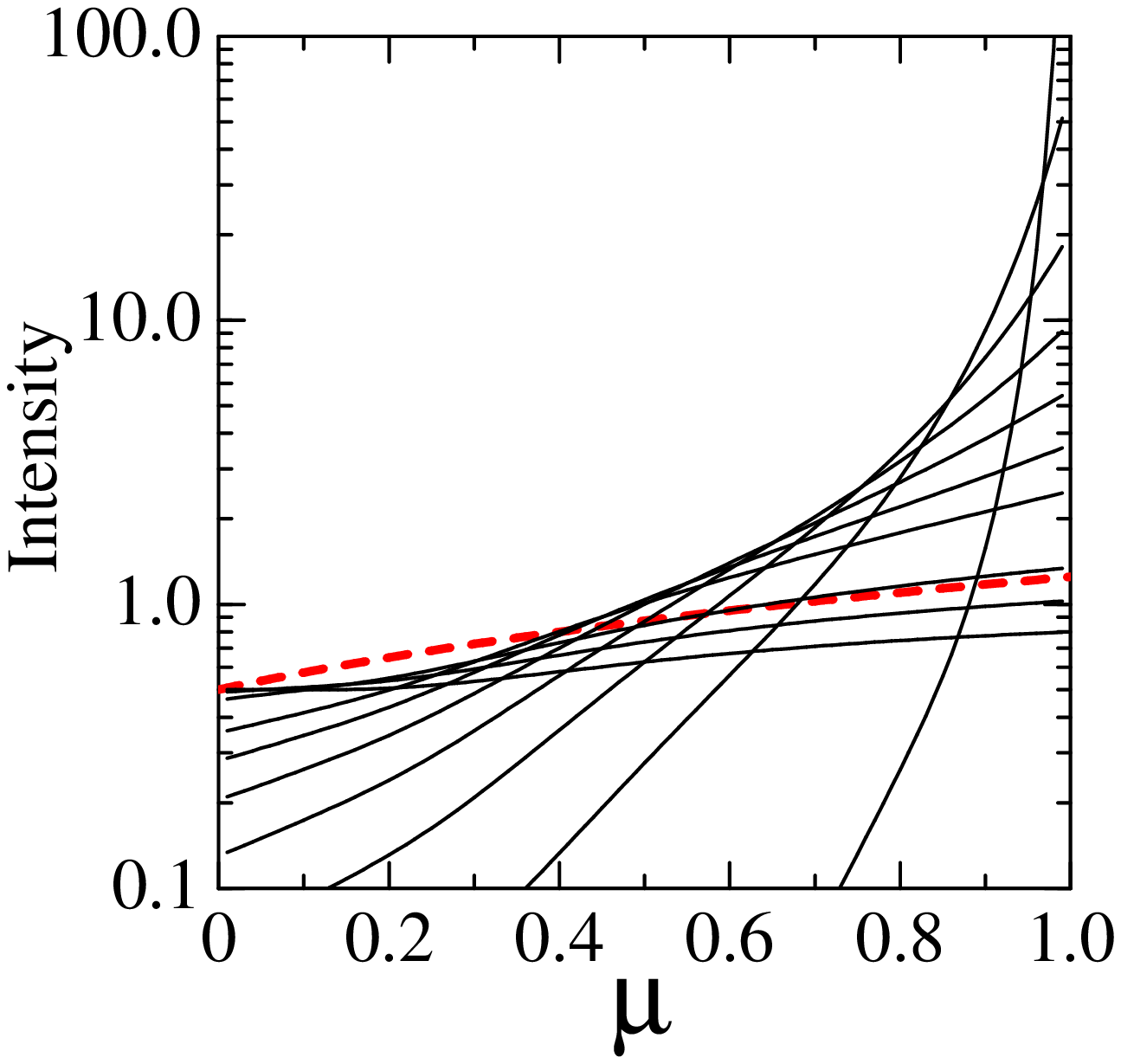}
  \end{center}
\caption{
Normalized emergent intensity
as a function of $\mu$
for several values of $\beta$:
(a) $\tau_0=0.1$ and (b) $\tau_0=1$
in the terminal case.
The values of $\beta$ are
0, 0.1, 0.2, 0.3, 0.4, 0.5, 0.6, 0.7, 0.8, 0.9, and 0.99:
as the flow speed becomes large, the intensity at $\mu=1$ is large.
The dashed line is for the usual
Milne-Eddinton solution for the plane-parallel case.
}
\end{figure}

In figure 1,
the emergent intensity $I(0, \mu, \beta)$
normalized by $I_{\rm s}$ is shown 
for several values of $\beta$ and $\tau_0$
as a function of $\mu$.

In the case of small $\tau_0$ (figure 1a),
when the flow speed is small,
the normalized emergent intensity is almost unity
except for small $\mu$ direction,
since the uniform source is seen except for small $\mu$ direction,
where the source function is seen.
When the flow speed becomes large, however,
the emergent intensity becomes remarkably anisotropic;
it decreases in the edgeward direction,
whereas it greatly increases in the poleward direction.
This is the {\it relativistic peaking effect},
which originates from the relativistic Doppler effect and aberration.

In the case of $\tau_0=1$ (figure 1b),
the emergent intensity for small $\beta$ is reduced
since the optical depth becomes large.
However, the relativistic peaking effect 
becomes effective as the disk optical depth becomes large.

\begin{figure}
  \begin{center}
  \FigureFile(80mm,80mm){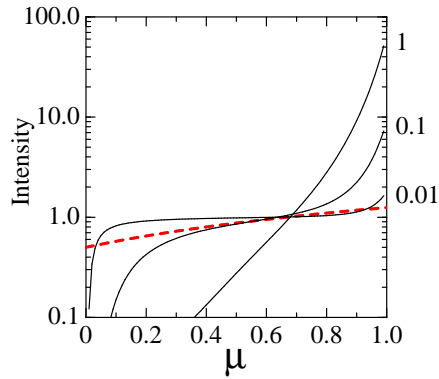}
  \end{center}
\caption{
Normalized emergent intensity
as a function of $\mu$
for several values of $\tau_0$
in the case of $\beta=0.9$
in the terminal case.
The values of $\tau_0$ are
0.01, 0.1, and 1.
The dashed line is for the usual
Milne-Eddinton solution for the plane-parallel case.
}
\end{figure}

In figure 2,
the normalized emergent intensity $I(0, \mu, \beta)$
is shown for several values of $\tau_0$
in the case of $\beta=0.9$.
As is seen in figure 2,
the relativistic peaking effect
strongly depends on the disk optical depth.

\subsection{Optically Thin Limit}

We next consider the flow
at a constant speed
using the closure relation (\ref{closure_pp})
in the optically thin limit.

Eliminating $E$ by the closure relation (\ref{closure_pp}),
the source functions (\ref{Sa_pp}) and (\ref{Ss_pp})
are expressed as
\begin{eqnarray}
   S_{\rm a} &=& 
           \frac{1}{4\pi}
           \frac{\kappa_0^{\rm abs} }{\kappa_0^{\rm abs}+\kappa_0^{\rm sca}}
           \frac{(1+\beta^2)cP - 2F\beta}{f-\beta^2}
\nonumber \\
   &=&
           \frac{1-A}{4\pi}
           \frac{(1+\beta^2)cP - 2F\beta}{f-\beta^2},
\label{Sa_pp2}
\\
   S_{\rm s} &=& 
           \frac{1}{4\pi}
           \frac{\kappa_0^{\rm sca} }{\kappa_0^{\rm abs}+\kappa_0^{\rm sca}}
           \frac{(1+\beta^2)cP - 2F\beta}{f-\beta^2}
           \frac{3}{4}
           \left[ 1+\frac{(\mu-\beta)^2}{(1-\beta \mu)^2}f
                   +\frac{(1-\beta^2)(1-\mu^2)}{(1-\beta \mu)^2}
                    \frac{1-f}{2}
           \right]
\nonumber \\
   &=&
           \frac{A}{4\pi}
           \frac{(1+\beta^2)cP - 2F\beta}{f-\beta^2}
           \frac{3}{4}
           \left[ 1+\frac{(\mu-\beta)^2}{(1-\beta \mu)^2}f
                   +\frac{(1-\beta^2)(1-\mu^2)}{(1-\beta \mu)^2}
                    \frac{1-f}{2}
           \right],
\label{Ss_pp2}
\end{eqnarray}
where
\begin{equation}
   A \equiv
           \frac{\kappa_0^{\rm sca} }{\kappa_0^{\rm abs}+\kappa_0^{\rm sca}}
\end{equation}
is the scattering albedo.

In the optically thin limit,
we can easily obtain the quantities of radiation fields
and the {\it velocity-dependent} Eddington factor
(cf. Fukue 2006b, 2007b; Koizumi and Umemura 2007).
That is to say,
if there exists a uniform source of intensity $I_0$
at the disk equator,
the radiation energy density $E$,
the radiative flux $F$, and the radiation stress tensor $P$
in the inertial (fixed) frame
are respectively calculated as
\begin{eqnarray}
   cE &=& 2\pi I_0,
\\
    F &=& \pi I_0,
\\
   cP &=& \frac{2}{3} \pi I_0.
\end{eqnarray}
Hence, from the closure relation (\ref{closure_pp})
the velocity-dependent variable Eddington factor $f(\beta)$
is derived as
\begin{equation}
   f(\beta) = \frac{1-3\beta+3\beta^2}{3-3\beta+\beta^2}.
\end{equation}

Using these expressions,
the source functions $S_{\rm a}$ and $S_{\rm s}$
are explicitly expressed as a function of $\beta$ and $\mu$:
\begin{eqnarray}
   S_{\rm a} &=& 
           \frac{1-A}{2} \frac{I_0}{3}
           \frac{3-3\beta+\beta^2}{1-\beta^2},
\label{Sa_pp3}
\\
   S_{\rm s} &=& 
           \frac{A}{2} \frac{I_0}{4}
           \frac{(1-\beta\mu)^2 (3-3\beta+\beta^2)
                 +(\mu-\beta)^2 (1-3\beta+\beta^2)
                 +(1-\beta^2)^2 (1-\mu^2)}
                {(1-\beta^2)(1-\beta\mu)^2}.
\label{Ss_pp3}
\end{eqnarray}
The emergent intensity (\ref{i_sol10}) does also
depend both on the flow speed and the direction cosine.
Examples of solutions are shown in figures 3 and 4.

\begin{figure}
  \begin{center}
  \FigureFile(80mm,80mm){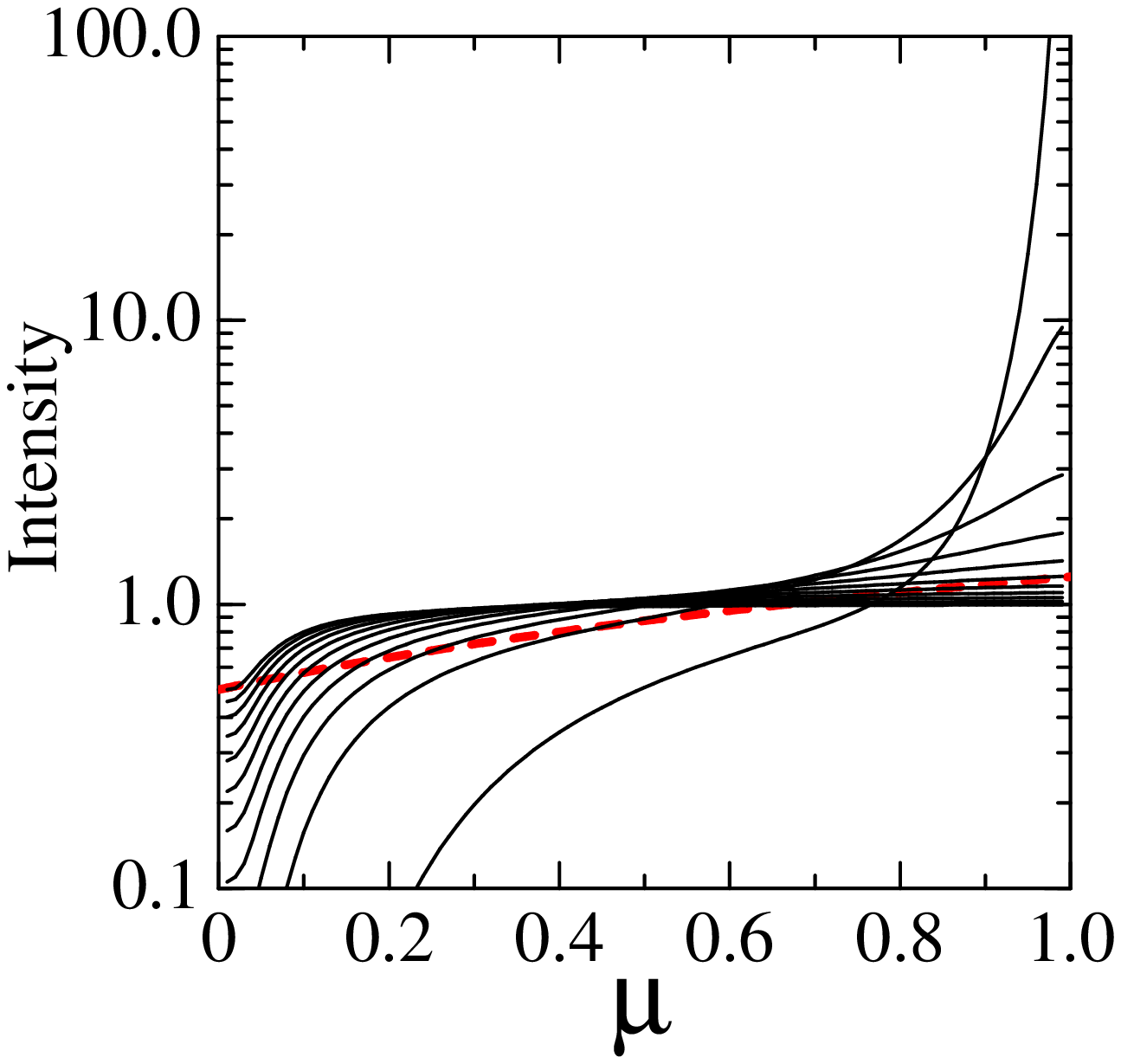}
  \FigureFile(80mm,80mm){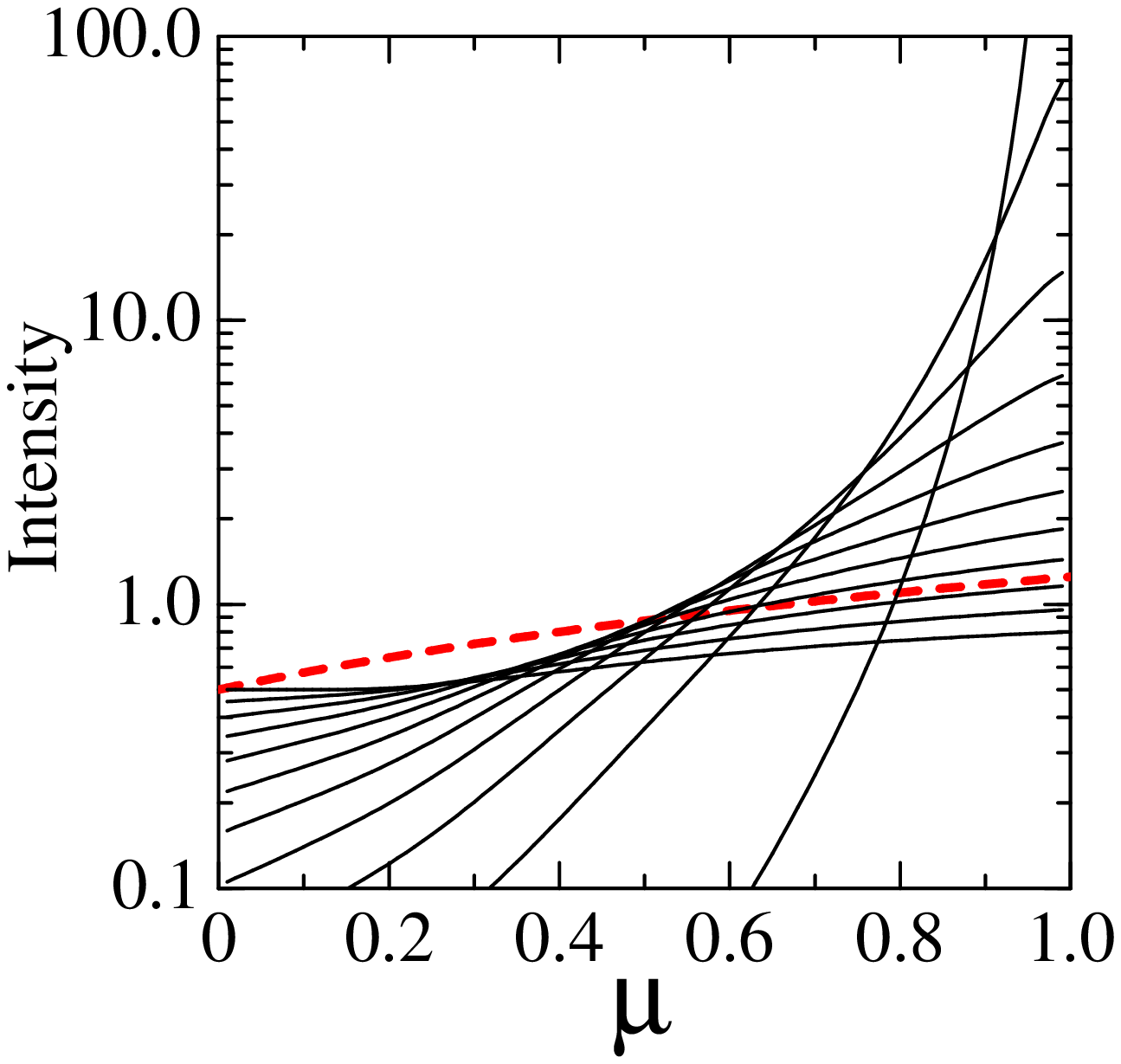}
  \end{center}
\caption{
Normalized emergent intensity
as a function of $\mu$
for several values of $\beta$:
(a) $\tau_0=0.1$ and (b) $\tau_0=1$
in the optically thin limit.
The values of $\beta$ are
0, 0.1, 0.2, 0.3, 0.4, 0.5, 0.6, 0.7, 0.8, 0.9, and 0.99:
as the flow speed becomes large, the intensity at $\mu=1$ is large.
The dashed line is for the usual
Milne-Eddinton solution for the plane-parallel case.
}
\end{figure}

In figure 3,
the emergent intensity $I(0, \mu, \beta)$
normalized by $I_0$ is shown 
for several values of $\beta$ and $\tau_0$
as a function of $\mu$.

The qualitative behavior of the emergent intensity
is similar to that of the terminal case.
That is, in the case of small $\tau_0$ (figure 3a),
when the flow speed is small,
the normalized emergent intensity is almost unity
except for small $\mu$ direction,
since the uniform source is seen except for small $\mu$ direction,
where the source function is seen.
When the flow speed becomes large, however,
the {\it relativistic peaking effect} becomes prominent.
In the case of $\tau_0=1$ (figure 3b),
the emergent intensity for small $\beta$ is reduced
since the optical depth becomes large.
However, the relativistic peaking effect 
becomes effective as the disk optical depth becomes large.

\begin{figure}
  \begin{center}
  \FigureFile(80mm,80mm){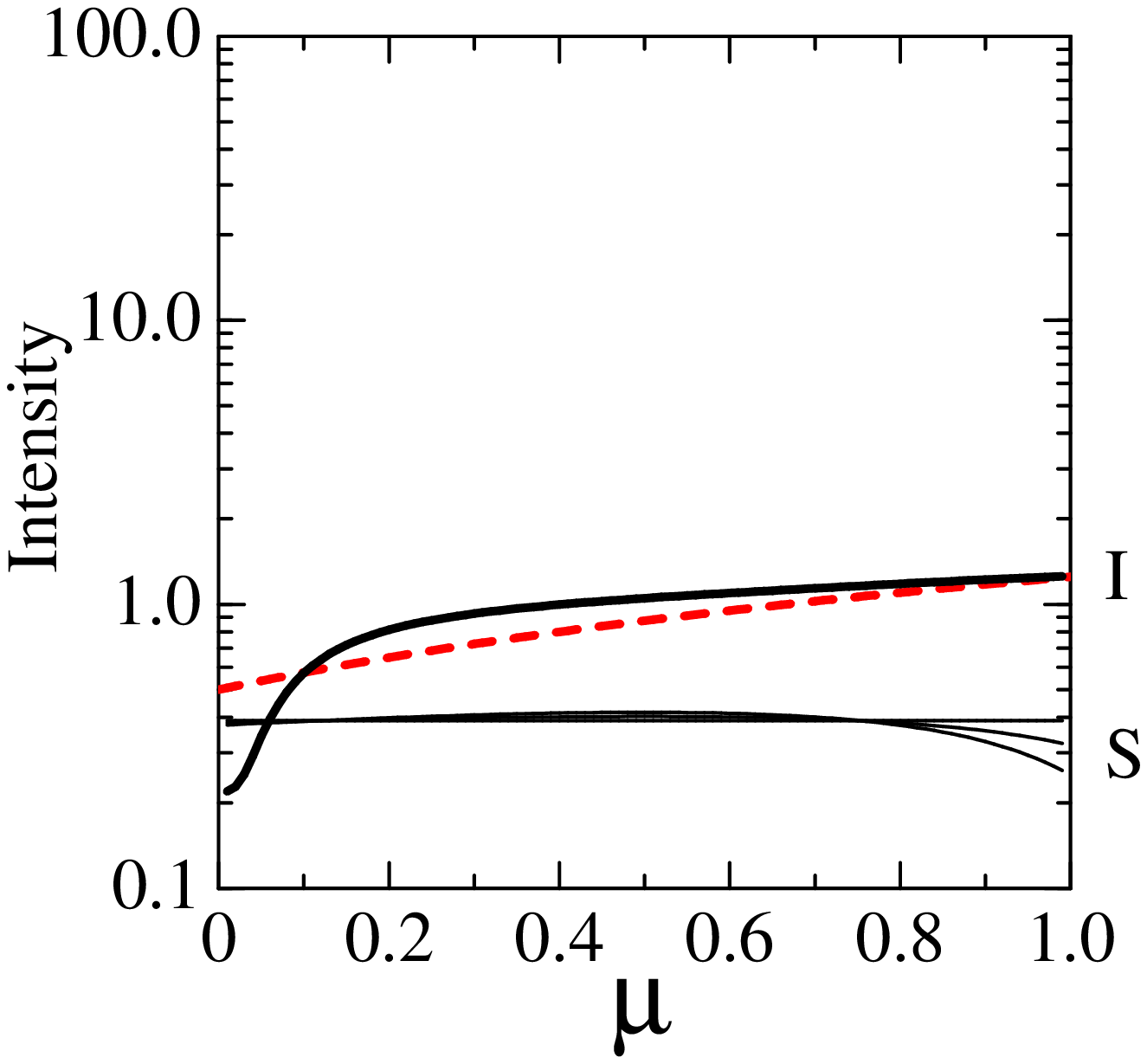}
  \FigureFile(80mm,80mm){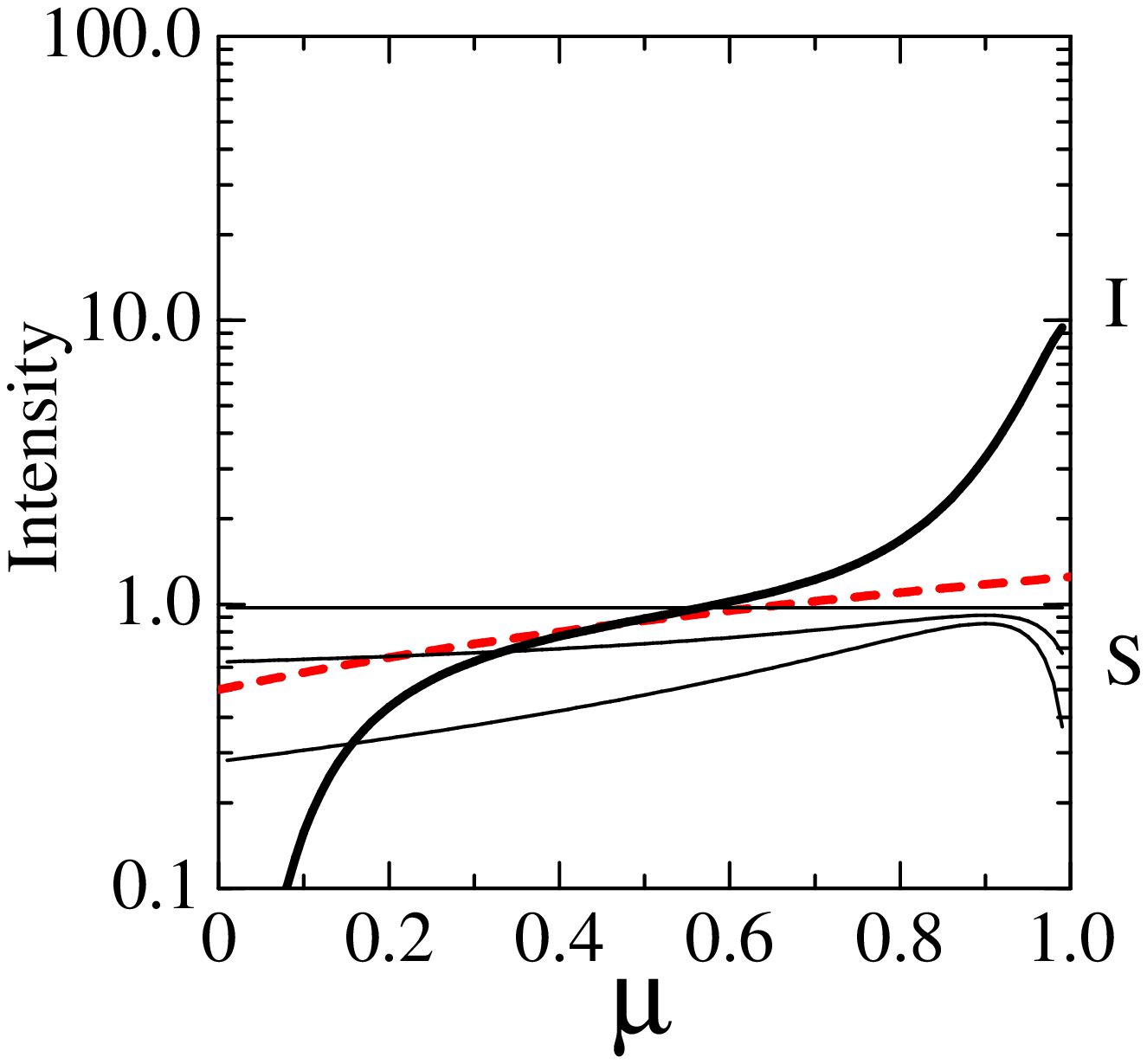}
  \end{center}
\caption{
Normalized emergent intensity and source functions
as a function of $\mu$
for several values of $\beta$
in the case of $\tau_0=0.1$
in the optically thin limit.
The values of $\beta$ are
(a) 0.5 and (b) 0.9.
Thin solid curves represent
$S_{\rm s}/A$, $S_{\rm a}+S_{\rm s}$ ($A=0.5$),
and $S_{\rm a}/(1-A)$, from bottom to top, respectively,
whereas
thick solid curves denote
$I(0, \mu, \beta)$ ($A=0.5$).
The dashed line is for the usual
Milne-Eddinton solution for the plane-parallel case.
}
\end{figure}

In figure 4,
the emergent intensity and the source functions
normalized by $I_0$ are shown
for several values of $\beta$.
In figure 4,
thin solid curves represent
$S_{\rm s}/A$, $S_{\rm a}+S_{\rm s}$ ($A=0.5$),
and $S_{\rm a}/(1-A)$, from bottom to top, respectively,
whereas
thick solid curves denote
$I(0, \mu, \beta)$ ($A=0.5$).

As is seen in figure 4 and equation (\ref{Sa_pp3}),
the source function $S_{\rm a}$
does depend only on the flow speed.
This is just because, in the present definition,
the source function $S_{\rm a}$
is proportional to the radiation energy density
$E_{\rm co}$ in the comoving frame,
since $cE_{\rm co} = cE - 2F\beta + \beta^2 cP$.
On the other hand,
the source function $S_{\rm s}$
depends on the direction cosine as well as the flow speed.
It is small both in the forward and backward directions,
while it becomes the maximum in the direction
at  $\mu = \beta$.

%%%%%%%%%%%%%  CONCLUDING REMARKS  %%%%%%%%%%%%%%%%%%%%%%%%%%%%

\section{Concluding Remarks}

In this paper 
we have examined the radiative transfer problem
in an accretion disk wind
under the plane-parallel approximation
in the fully special relativistic regime.
For an equilibrium flow,
where the radiative quantities and source function are constant,
we analytically obtain the specific and emergent intensities.
We found that the emergent intensity depends
on the flow speed as well as the direction cosine,
and exhibits a relativistic peaking effect.
As a result,
a wind luminosity would be overestimated by a pole-on observer
and underestimated by an edge-on observer,
when we observe an accretion disk wind
(cf. Sumitomo et al. 2007; Nishiyama et al. 2007).

It should be noted that
the {\it apparent} optical depth
in the relativistically moving media.
Abramowicz et al. (1991) pointed out that
the optical depth in the relativistic flow
decreases as $\gamma(1-\beta\mu)\tau$ toward the downstream direction,
due to the Doppler and aberration effects.
Inspecting equation (\ref{itransf_pp3}) or solution (\ref{i_sol13}),
we find that in the present case
the optical depth $\tau$ is apparently replaced by
$\gamma (1-\beta\mu)\tau$.
This is just consistent with the results
by Abramowicz et al. (1991).

In this paper we only examined the frequency-integrated intensity
under the plane-parallel and gray approximations.
It should be briefly remarked on the frequency dependence;
i.e., the frequency-dependent intensity $I_\nu$.
As long as the opacity is gray,
the spectral transformation is determined
by the relativistic effect in the present situation.
Namely, for the frequency-integrated intensity
the relativistic invariant is $I/\nu^4$,
and this effect appears in solution (\ref{i_sol13})
and other equations as a factor of $[\gamma(1-\beta\mu)]^4$.
Since the relativistic invariant for the frequency-dependent intensity
is $I_\nu/\nu^3$,
the corresponding factor should be changed as
$[\gamma(1-\beta\mu)]^3$.
On the other hand, 
the relativistic modification in the optical depth discussed above
is not changed in the frequency-dependent case.
Hence, the present results may be valid
for the frequency-dependent emergent intensity;
e.g., the incident intensity $I_{\nu 0}$
would be boosted in the polar direction,
according to solution (\ref{i_sol13}),
but $[\gamma(1-\beta\mu)]^4$ is replaced by $[\gamma(1-\beta\mu)]^3$.

The present study can be applied to energetic jets in,
e.g., gamma-ray blazars and gamma-ray bursts.
The effect of the relativistic jets
on the emergent spectrum has been studied in several literatures
  (Dermer, Schlickeiser 1993, 2002; Dermer 1998;
  B\"ottcher, Dermer 2002; Dermer et al. 2007),
In these earlier works, however,
the radiation transfer was not treated at all.
Hence, the present approach may be very usefull
in these fields of active phenomena.

The radiative transfer problem investigated in the present paper
must be quite {\it fundamental problems} for
accretion disk physics and astrophysical jet formation.
In onder to demonstrate the existence of the relativistic peaking effect,
we have imposed various assumptions,
including a constant flow speed, gray approximation,
no heating source, and so on.
By relaxing these assumptions
and integrating the relativistic transfer equation numerically,
we could obtain the emergent intensity and spectra
more quantitatively.
These are left as future works.

\vspace*{1pc}

The author would like to thank
S. Kato, S. Mineshige, M. Umemura, T. Koizumi, and C. Akizuki
for enlightening and stimulating discussions.
%The authors would like to thank Professor Y. Osaki and Dr. N. Shibazaki
%for their valuable comments and discussions.
%The author would like to thank an anonymous referee for valuable comments.
This work has been supported in part
by a Grant-in-Aid for Scientific Research (18540240 J.F.) 
of the Ministry of Education, Culture, Sports, Science and Technology.

%\noindent

\end{document}